\documentclass{aa}

\usepackage{graphicx}
\usepackage{color}
\usepackage{natbib}
\usepackage{txfonts}

\def\apj{ApJ} 
\def\apjl{ApJL} 
\def\aap{A\&A} 
\def\aaps{A\&ASS}

\def\solphys{Sol~.Phys.}
\newcommand{\CaII}{\ion{Ca}{ii}}

\begin{document}

%
%

\title{On the fine structure of the quiet solar \CaII~K atmosphere}

\author{A. Tritschler\inst{1,2,3}, W. Schmidt\inst{2}, H. Uitenbroek\inst{3}, and S. Wedemeyer-B\"ohm\inst{2}}
   
\institute{Big Bear Solar Observatory, New Jersey Institute of Technology, 
           40386 North Shore Lane, \\ Big Bear City, CA-92314, U.S.\\
           \email{ali@bbso.njit.edu, ali@kis.uni-freiburg.de, ali@nso.edu}
           \and Kiepenheuer--Institut f\"ur Sonnenphysik, Sch\"oneckstr. 6, D--79104 Freiburg, Germany\\
           \email{wolfgang@kis.uni-freiburg.de, wedemeyer@kis.uni-freiburg.de}
           \and National Solar Observatory/Sacramento Peak\thanks{Operated by the 
                Association of Universities for Research in Astronomy, Inc. (AURA), 
                for the National Science Foundation}, P.O.~Box 62, Sunspot, NM-88349, U.S.A.\\
           \email{huitenbroek@nso.edu}}
     
\titlerunning{Fine structure of the \CaII~K atmosphere}

\authorrunning{Tritschler et al.}

\offprints{A. Tritschler}

\date{\today}

%
%
\abstract
{}
{We investigate the morphological, dynamical, and evolutionary properties of the 
          internetwork and network fine structure of the quiet sun at disk centre.}
          {The analysis is based on a $\sim$6\,h time sequence of 
          narrow-band filtergrams centred on the inner-wing
          \CaII~K$_{\rm 2v}$ reversal at 393.3\,nm.
          To examine the temporal evolution of network and internetwork areas separately 
          we employ a double-Gaussian decomposition of the mean intensity distribution.
          An autocorrelation analysis is performed to determine 
          the respective characteristic time scales. In order to analyse statistical 
          properties of the fine structure we apply image segmentation techniques.}
          {
          The results for the internetwork are related to predictions derived from 
          numerical simulations of the quiet sun. 
          The average evolutionary time scale of the internetwork in our observations
          is 52\,sec. Internetwork grains show a tendency to appear 
          on a mesh-like pattern with a mean cell size
          of $\sim$4-5\,arcsec. Based on this size and the spatial organisation of the mesh
          we speculate that this pattern is related 
          to the existence of photospheric downdrafts 
          as predicted by convection simulations. 
          The image segmentation shows that typical sizes of both network and 
          internetwork grains are in the order of 1.6\,arcs.
         }
         {}
\maketitle

%
%

\section{Introduction}

The solar chromosphere is the highly structured and very dynamic
atmospheric link between the photosphere and the thin transition layer to the hot
corona. Understanding its physical
structure is most likely to shed light on outer-atmosphere
heating mechanisms and the acceleration and composition of
the solar wind. Well known chromospheric diagnostics are
the cores and inner wings of strong absorption lines like H$\alpha$
and \CaII~H \& K. 
Since the early work of \citet{jensen+orrall1963}, 
\citet{orrall1966}, and \citet{cram1978} decades ago, the
resonance lines have been extensively employed to explore chromospheric
dynamics and chromospheric structure. For reviews and references 
therein we refer to \citet{judge+peter1998} and \citet{rutten1999}.

In low-spatial resolution filtergrams the internetwork regions appear
rather homogeneous and dark, while the highly structured and inhomogeneous 
face of the chromosphere in the internetwork is only revealed in 
high-resolution filtergrams. Internetwork areas are intermittently filled with
brightenings. These internetwork brightenings are most conspicuous in
the emission feature in the blue wing of the \CaII~K and H
lines. Thus, they have been named \CaII~K(H)$_{\rm 2v}$ grains or simply
{\it internetwork grains}. The investigation of the properties of these grains is a
diagnostic tool to probe chromospheric conditions in internetwork
regions. For a detailed description of the K$_{\rm 2v}$ grain phenomenon
we refer to \cite{rutten+uitenbroek1991} and references therein.
%
\begin{figure*}[th]
  \centering
  \includegraphics[width=\textwidth]{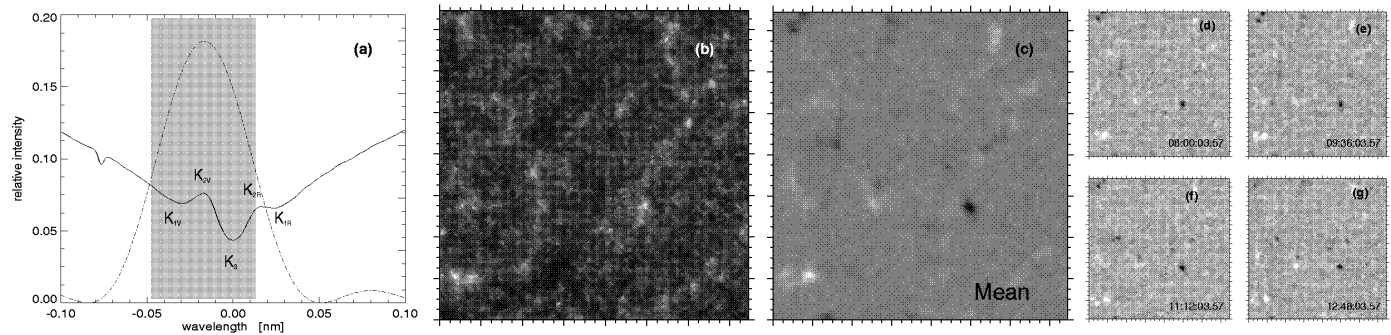}
  \caption{(a): Theoretical transmission curve of the Lyot filter (dot-dashed)
           and the mean spectrum of the \CaII~K line taken from a photospheric spectral
           atlas \citep{neckel1999}. The shaded area indicates the 60\,pm FWHM range.
           (b): Best filtergram taken in the \CaII~K line.
           One minor tick corresponds to 5\,arcsec. The filtergram is displayed on a logarithmic scale.
           (c)-(g): Mean (large) and individual (small) magnetograms taken from MDI.
           One minor tick corresponds to 5\,arcsec.}
  \label{fig:obs_sample_mdi}
\end{figure*}
%

Where does the intermittent chromospheric structure in space and time originate from?
Observational evidence makes clear that the dynamic behaviour and radiative
signature of the chromosphere is determined by its response to upward travelling waves
\citep{wikstol+etal2000, judge+tarbell+wilhelm2001} excited by the effects of 
convection \citep{hoekzema+rutten1998, skartlien+stein+nordlund2000, 
hoekzema+rimmele+rutten2002}. While the waves propagate 
upwards they steepen into shocks and form  high-frequency shock sequences 
pervading the internetwork regions.
The formation of shocks is strongly supported by the 
numerical simulations of \citet{carlsson+stein1992a, carlsson+stein1992b, 
carlsson+stein1997}. However, because of the 
one-dimensionality of these simulations, interference with 
slanted waves cannot be taken into account and shock 
trajectories are assumed to be radial. 
Recent three-dimensional radiative hydrodynamic simulations
result in a dynamic and thermally bimodal structure of 
the non-magnetic chromospheric layers \citep{wedemeyer+etal2004}. 
The simulations show that upward travelling and interfering waves and shocks 
lead to the formation of a network of hot upwelling filaments and enclosed cool regions. 
The non-magnetic numerical modelling corroborates observational findings
that the internetwork atmosphere and its behaviour are predominantly
determined by the chromospheric response to acoustic oscillations. 

Magnetic fields play a minor role in the internetwork grain generation 
\citep{uexkuell+kneer1995, remling+deubner+steffens1996, hofmann+steffens+deubner1996, 
steffens+hofmann+deubner1996, hoekzema+rutten1998, lites+rutten+berger1999, 
pontieu2002, hoekzema+rimmele+rutten2002} and are 
only occasionally involved in form of the
{\it flashers} \citep{brandt+etal1992, nindos+zirin1998, krijger+etal2001, dewijn+etal2005}. 
However, magnetic fields appear to play a major role in the higher atmosphere
where plasma $\beta$ (ratio of gas pressure over magnetic pressure) becomes small
\citep{mcintosh+etal2001} and wave reflection and conversion 
between acoustic and magneto-acoustic 
modes takes place \citep{rosenthal+etal2002, bogdan+etal2003}.

In the present work we investigate
the \CaII~K chromosphere and its distinct fine structure based on an analysis of a
long time sequence ($\sim$6\,h) of narrow-band (60\,pm FWHM) filtergrams featuring
a spatial resolution of $0.7$\,arcsec. We concentrate on 
morphological, dynamical, and evolutionary 
differences between the network and the internetwork. We employ
image segmentation techniques to establish statistics of properties
like size and nearest-neighbour distance that can be used to validate current and future 
simulations. 

The paper is organised as follows. The observations are introduced and 
the data reduction process is reviewed in Sect.~\ref{sec:obs} and 
Sect.~\ref{sec:data}, respectively. In Sect.~\ref{sec:imgseg} 
we go into image segmentation processing and describe the applied
techniques (a) to distinguish on large-scales between the 
network and the internetwork, and (b) to identify and isolate 
small-scale chromospheric brightenings. 
In Sect.~\ref{sec:temp_evol} the temporal evolution is studied 
as a result from a correlation analysis 
Section \ref{sec:statistics} 
is dedicated to a morphological analysis of chromospheric brightenings.
The results of our investigation are concluded in Sect.~\ref{sec:conclusions}.

%
%

\section{Observations}\label{sec:obs}

The observations discussed here have been carried out at the Vacuum Tower Telescope of
the Kiepenheuer-Institut at the Observatorio del Teide on Tenerife, May 7, 1999. 
A Halle Lyot filter was used to obtain
narrow-band filtergrams in the \CaII~K line at 393.3\,nm. The
width of the filter was set to 60\,pm
and the transmission band was centred $\sim$0.017\,nm away from line core on the
K$_{\rm 2v}$ emission reversal. A $\sim$6\,h sequence was recorded with a cadence of 6.04\,s 
on a blue-sensitive 1024$\times$1024\,pixel CCD camera operated 
in a 512$\times$512 summing mode. The resulting image scale on the detector
was 0.35$\times$0.35\,arcsec$^2$. The exposure time was
250\,msec. For image stabilisation the VTT correlation tracker system
was used \citep{schmidt+kentischer1995, ballesteros+etal1996}.
%
\begin{figure}[t]
  \centering
  \includegraphics[width=8cm]{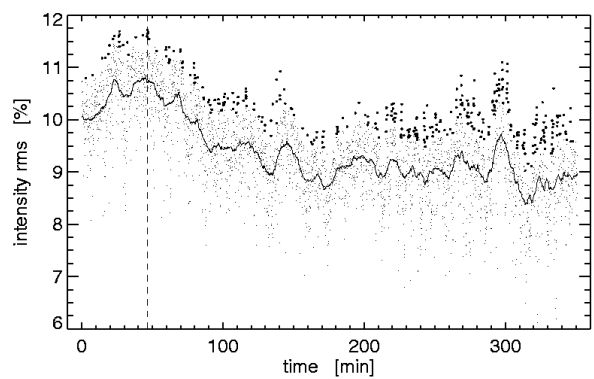}
  \caption{Rms-intensity fluctuation as a function of time 
           during the observation (small dots). Solid: smoothed version of
           the fluctuation used to determine all points that are $>$7.5\,\%
           from the local average value (large dots). The vertical dashed line indicates
           the time where the best filtergram displayed 
           in Fig.~\ref{fig:obs_sample_mdi} was taken.}
  \label{fig:rms07}
\end{figure}
%
Figure \ref{fig:obs_sample_mdi} (a) shows the atlas
profile  of the central part of the \CaII~K line taken from
\citet{neckel1999} and the theoretical transmission function of the
Lyot filter. Figure \ref{fig:obs_sample_mdi} (b) shows the best \CaII~K filtergram of the
observed quiet sun region. The observed intensities in the
filtergram are dominated by the K$_{\rm 2v}$ emission feature and
contaminations from K$_3$ and K$_{\rm 1v}$ are strongly suppressed due
to the rather sharp decline of the transmission function. Even in
case the K$_{\rm 2v}$ reversal is absent, the main contribution to
the integrated intensity comes from the range of heights that contribute
to reversal features, although at a much reduced source function. The
influence from red-shifted K$_{\rm 1v}$ or blue-shifted K$_3$ is
expected to be small within the 60\,pm filter passband in our
observations. To estimate the formation height range more quantitatively 
we performed a preliminary analysis employing response functions to examine the influence of
temperature changes in the solar atmosphere on the filter integrated
signal \citep[see ][]{fossum+carlsson2005}. From the response 
functions computed through a hydrostatic model of the solar atmosphere 
we conclude that our filtergram observations basically 
reflect conditions prevailing in the height range 350-650\,km.
Experiments show that this height range is very similar 
in one-dimensional snapshots from simulations of 
chromospheric dynamics \citep[e.g. ][]{carlsson+stein1997}. 
For a more thorough study of the influence of
temperature changes in the solar atmosphere on the response of the
filter integrated signal we refer to \cite{uitenbroek2006}.
%
\begin{figure*}[t]
  \centering
  \includegraphics[width=\textwidth]{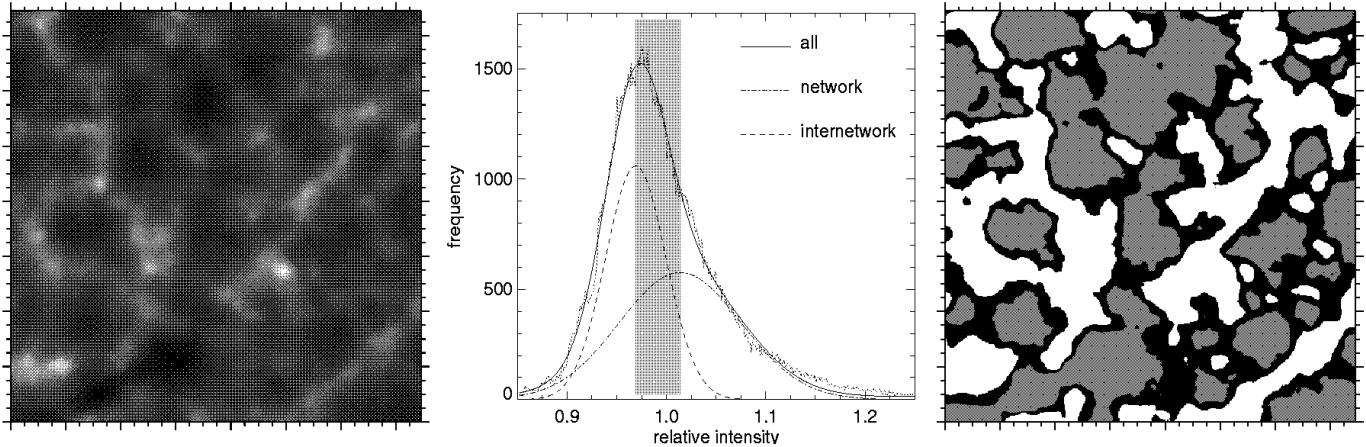}
  \caption{Left: Mean intensity map (averaged over the whole
           sequence). Middle: histogram of the mean intensity map
           (dotted), double-Gaussian fit to the data points (full) 
           and single-Gaussian profiles (dashed, dashed-dotted). 
           The shaded area indicates the range of relative intensities that 
           is excluded from the statistical analysis.
           Right: Masks used for the definition of the internetwork (grey) 
           and the network (white). Black areas define regions which
           do neither belong to the network nor to the 
           internetwork according to our definition.}
  \label{fig:avg_hist_mask}
\end{figure*}
%

MDI on-board SoHO acquired magnetograms and
continuum intensity data in the full-disk mode (2\,arcsec spatial
resolution) with a cadence of 96\,min. These data were averaged in
time, rescaled and co-aligned to the mean \CaII~K filtergram
averaged over the whole $\sim$6\,h observing period
(Fig.~\ref{fig:avg_hist_mask}, left). Figure \ref{fig:obs_sample_mdi}
(c) shows the resulting mean magnetogram and the four
frames (d-g) taken during our
observing period. The magnetograms show several magnetic elements up
to the size of $\sim$10\,arcsec. These structures are persistent
during the whole observation and coincide spatially with bright
regions in \CaII~K outlining the chromospheric network (see
Fig.~\ref{fig:obs_sample_mdi}b). The smallest structures
visible in our data in the internetwork regions
span 2-3\,pixels corresponding to 0.7-1.0\,arcsec. This puts an 
upper limit on the spatial resolution.  

%
%

\section{Data reduction}\label{sec:data}

All data were corrected for dark current and offset and for intensity
inhomogeneities caused by the CCD or any optical surface. The guiding
was not perfect throughout the whole sequence; occasionally, the
correlation tracker, working on normal granulation, lost its tracking
position in moments of bad seeing. Therefore the images of the
series were aligned by shifting and minimising the root-mean-square difference 
with a precision of about 0.2\,pixel. To
this end, every 100th image was selected as a reference
image, and these images were aligned first. If the chosen
reference image had poor contrast, we used a
neighbouring, good-quality image instead. In a second step, the full
sequence was then aligned with respect to the nearest reference
image. The post-facto alignment procedure reduced the common field of view to about
150$\times$150\,arcsec$^2$. The remaining image displacements are
below 1\,pixel, and are  mostly due to distortion within the field of
view. They do not affect our data analysis. 

To characterise the quality of our data set we computed
the root-mean-square (rms) intensity contrast for each image, 
which is displayed in Fig.~\ref{fig:rms07}. The rms values 
fluctuate around a mean of $\sim$9\,\%, with a maximum and minimum 
of almost 12\,\% and 6\,\%, reflecting the best and the worst 
moments, respectively. 

%
%

\section{Image segmentation}\label{sec:imgseg}

This section introduces the image segmentation methods
we employ to determine statistical properties 
of internetwork and network fine structure (see Sect.~\ref{sec:statistics}). 
To distinguish between internetwork and network areas on 
large spatial scales we employ a threshold technique,
and to identify and isolate individual bright
structures on relatively small spatial scales, 
we apply morphological image processing
techniques.

\subsection{Definition of network and internetwork}\label{ssec:masks}

We use the very different temporal behaviour
of the network and the internetwork to distinguish between the two. 
After normalising the intensities of each individual
filtergram to the mean intensity of the image $I_{\rm mean}$
the mean intensity map (Fig.~\ref{fig:avg_hist_mask}, left) 
was calculated by averaging over all 3483 single
filtergrams. Motivated by the visual impression that the intensity
map is dominated by two different populations of intensity
we fitted the histogram (Fig.~\ref{fig:avg_hist_mask}, middle) of 
the mean intensity map by a double-Gaussian function. 
The two maxima of the Gaussian
components, located at 0.97 and 1.02, were used as intensity thresholds: pixels darker
than $<0.97$ and brighter than $>1.02$ are considered as internetwork
and network, respectively \citep[see e.g. ][]{kneer+uexkuell1993, steffens+hofmann+deubner1996, 
krijger+etal2001}. The right panel of Fig.~\ref{fig:avg_hist_mask} 
shows the binary mask that results from
thresholding the mean intensity map with the values provided by the
double-Gaussian fit. White areas define network regions and grey areas
define internetwork regions. Black areas are indefinite and are not
included in the analysis described in Sect.~\ref{sec:temp_evol} and
Sect.~\ref{sec:statistics}. If not mentioned otherwise 
in the following all intensities are 
given in units of $I_{\rm mean}$.

\subsection{Morphological image processing}\label{ssec:morph}

We pursue an image segmentation (or tessellation) approach to identify
and isolate individual chromospheric brightenings in the network and
the internetwork. We use an IDL implementation of the watershed
transformation to segment each filtergram into watershed regions
(cells) and their boundaries. As a result each individual cell
harbours one maximum and thus isolates and defines a single
structure. The transformation makes use of the morphological watershed
operator labelling each watershed region with an unique index, and
boundaries set to zero. For further details see \citet{dougherty1992}. 
Prior to the image segmentation process an optimum filter is applied 
to each filtergram to suppress high-frequency noise 
and to avoid over-segmentation.

\begin{figure}[t]
  \centering
  \includegraphics[width=4.25cm]{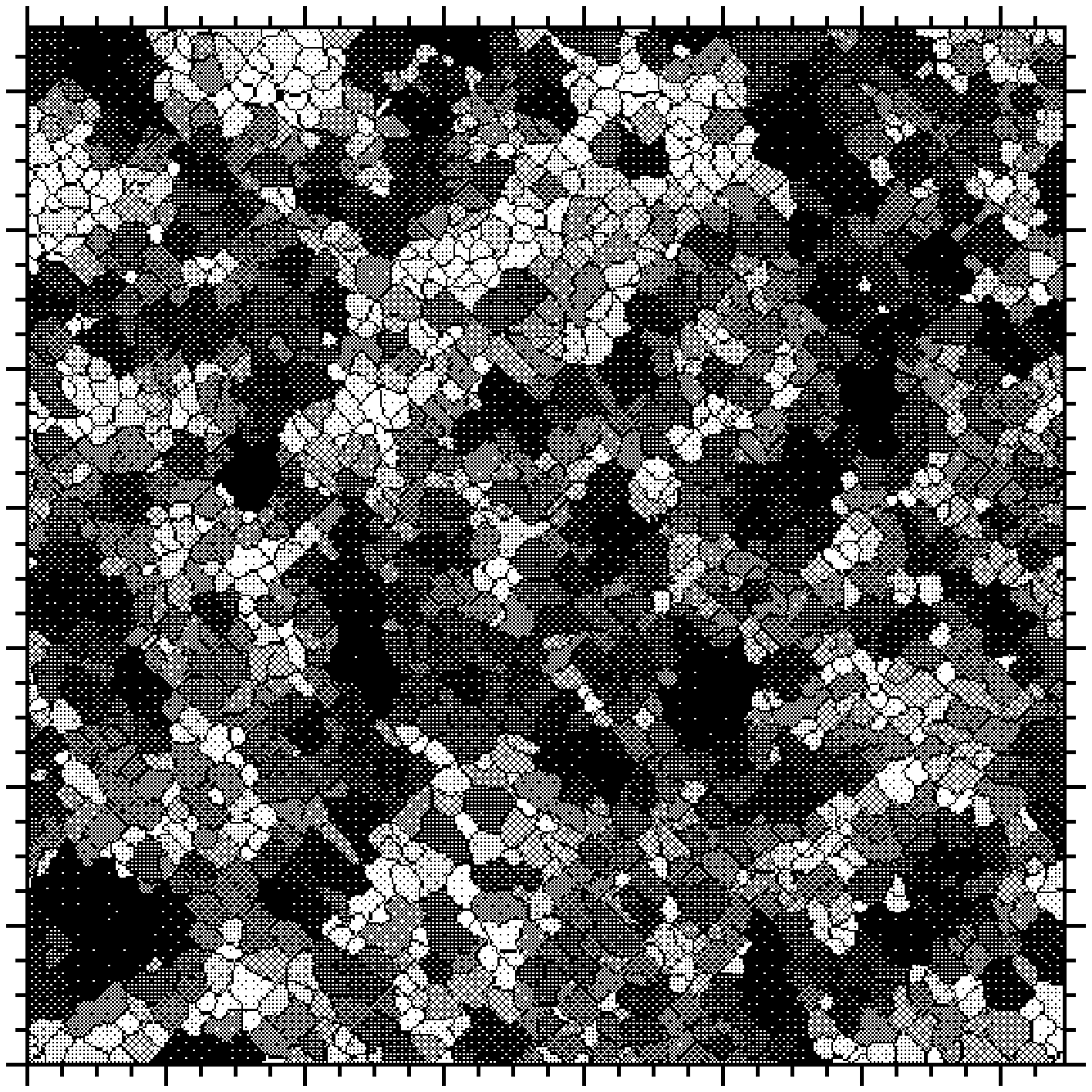}
  \includegraphics[width=4.25cm]{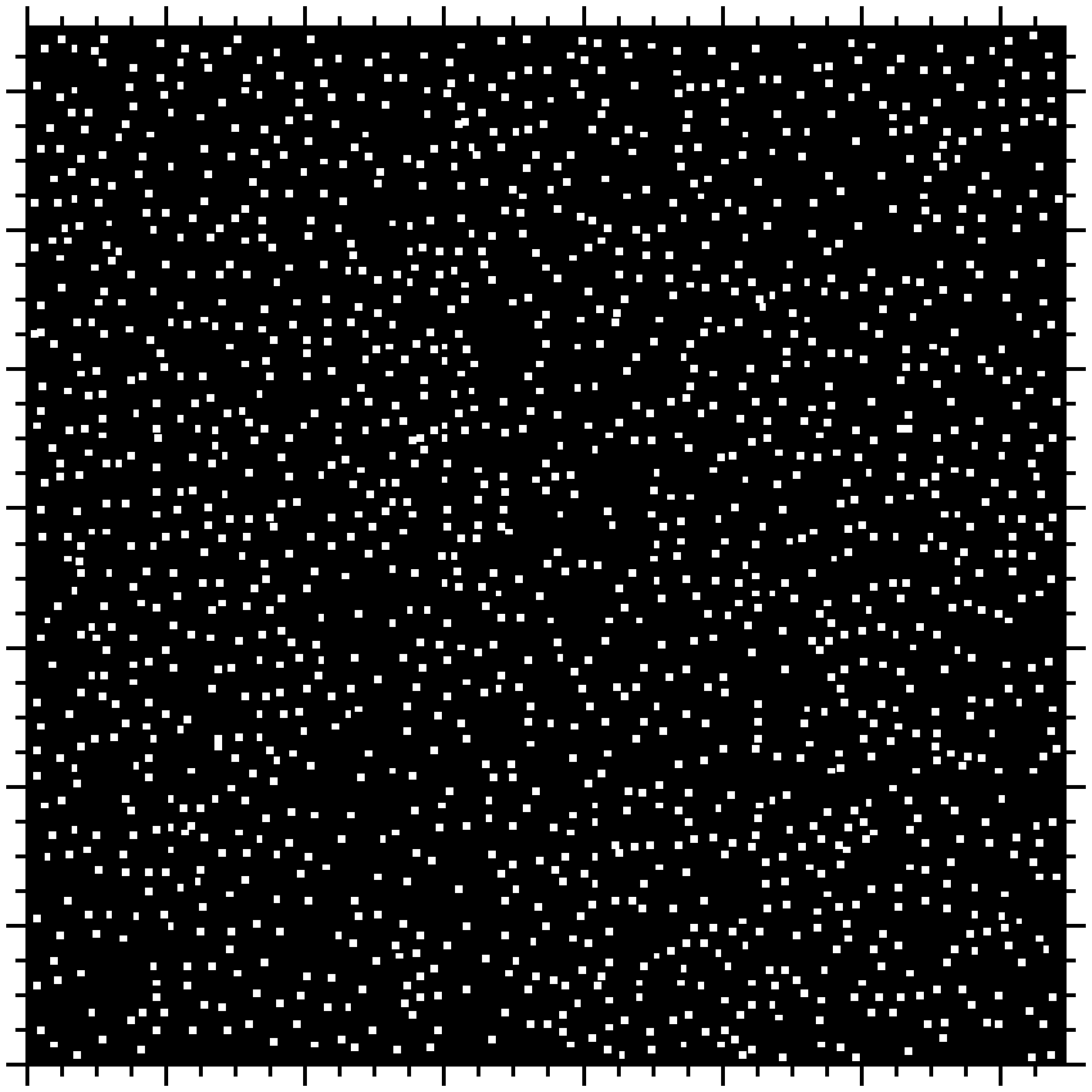}
  \caption{Left: Segmented image as a result from the watershed transformation.
           Right: Binary mask indicating the intensity peak position of identified 
           fine structure from the watershed transformation. 
           For better visibility the positions 
           are marked using more than one pixel.}
  \label{fig:segment}
\end{figure}

Figure \ref{fig:segment} (left) visualises the tessellation 
for a sample filtergram. We estimated the background intensity 
of each identified structure inside the cell by averaging over the cell boundary contours. The number of cell
pixels define the cell size and the average over these pixels defines the mean cell
intensity. From the mean background intensity and the peak
intensity we derive the full-width-half-maximum size
(effective diameter) of the corresponding structure \citep[see ][]{tritschler+schmidt2002b}.
Finally, each of the identified maxima are connected to its
neighbouring maxima via a mathematical triangulation which
returns an adjacency list from which nearest-neighbour distances can
be computed directly.
%
%

In addition we determine all local maxima along each
intensity row and column via numerical
differentiation. Local maxima in the image are then found by 
identifying all the pixels where local maxima in row and column coincide.
A comparison shows that these local maxima match very well 
with those found using the watershed transformation. 
Although the peak intensity is determined by the image segmentation
process we do not evaluate on the statistics of those but use it only for 
thresholding purposes.

%
%

\section{Temporal evolution of the chromosphere}\label{sec:temp_evol}

In the following, we analyse the temporal evolution of the
chromospheric network and internetwork regions. 
First, we deduce characteristic time scales
for the network and internetwork, separately. Second, the results from
the segmentation processing are used to visualise and study the
appearance of the fine structure.
Third, we examine the temporal and spatial frequency
behaviour using the $k-\nu$-diagram as 
well as power maps as a diagnostic tool.
%
\begin{figure}[tp]
  \centering
  \includegraphics[width=8.5cm]{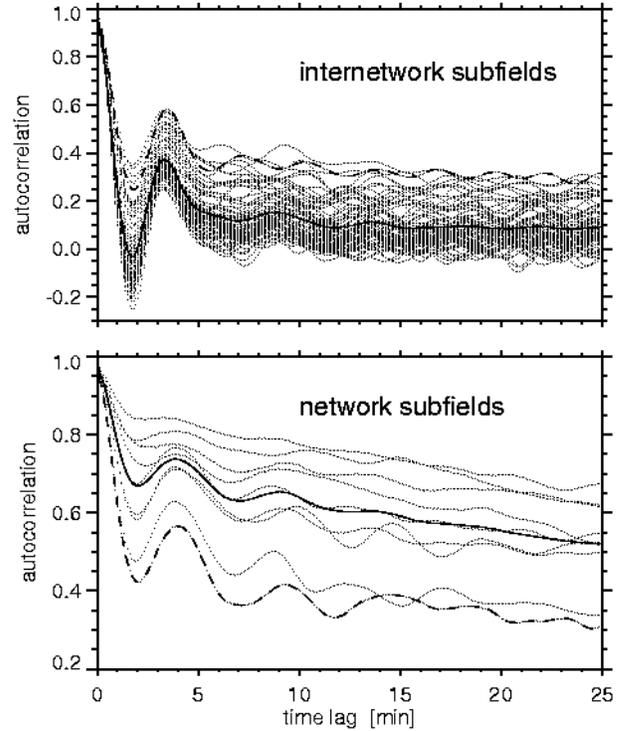}
  \caption{Autocorrelation versus time. Dots: individual autocorrelation
           functions for all subfields that belong either
           to the internetwork (upper panel, 73 subfields corresponding to the 
           filled circles in Fig.~\ref{fig:time_scales}, right) or to
           the network (lower panel, 10 subfields corresponding to the open 
           circles in Fig.~\ref{fig:time_scales}, right).
           Thick solid: average over all network/internetwork subfields.
           Thick dash-dotted: individual autocorrelation functions.}
  \label{fig:rho}
\end{figure}

\subsection{Autocorrelation Analysis}\label{ssec:time_scales}

In order to specify the temporal evolution in more detail
characteristic time scales are derived from an autocorrelation
analysis. To compare with the numerical results of \citet{wedemeyer+etal2004}
the whole field-of-view is divided into 22$\times$22\,pixel
subfields with 50\,\% overlap so that the subfield size corresponds to
the simulation size which covers a $\sim$5.6$\times$5.6\,Mm$^2$ region of the quiet Sun horizontally. 
Each individual subfield is allocated to the network or to
the internetwork if 100\,\% of the pixels coincide with network or
internetwork pixels as defined by a mask similar to the one described
in Sect.~\ref{ssec:masks} but using an intensity map averaged over
$\sim$75\,min (instead of the full $\sim$350\,min). For
computational reasons we performed the autocorrelation analysis for all subfields
on the first 750 images ($\sim$75\,min) only which provides a statistical relevance of
the autocorrelation function up to 25\,min.
%
\begin{figure}[tp]
  \centering
  \includegraphics[width=4.75cm]{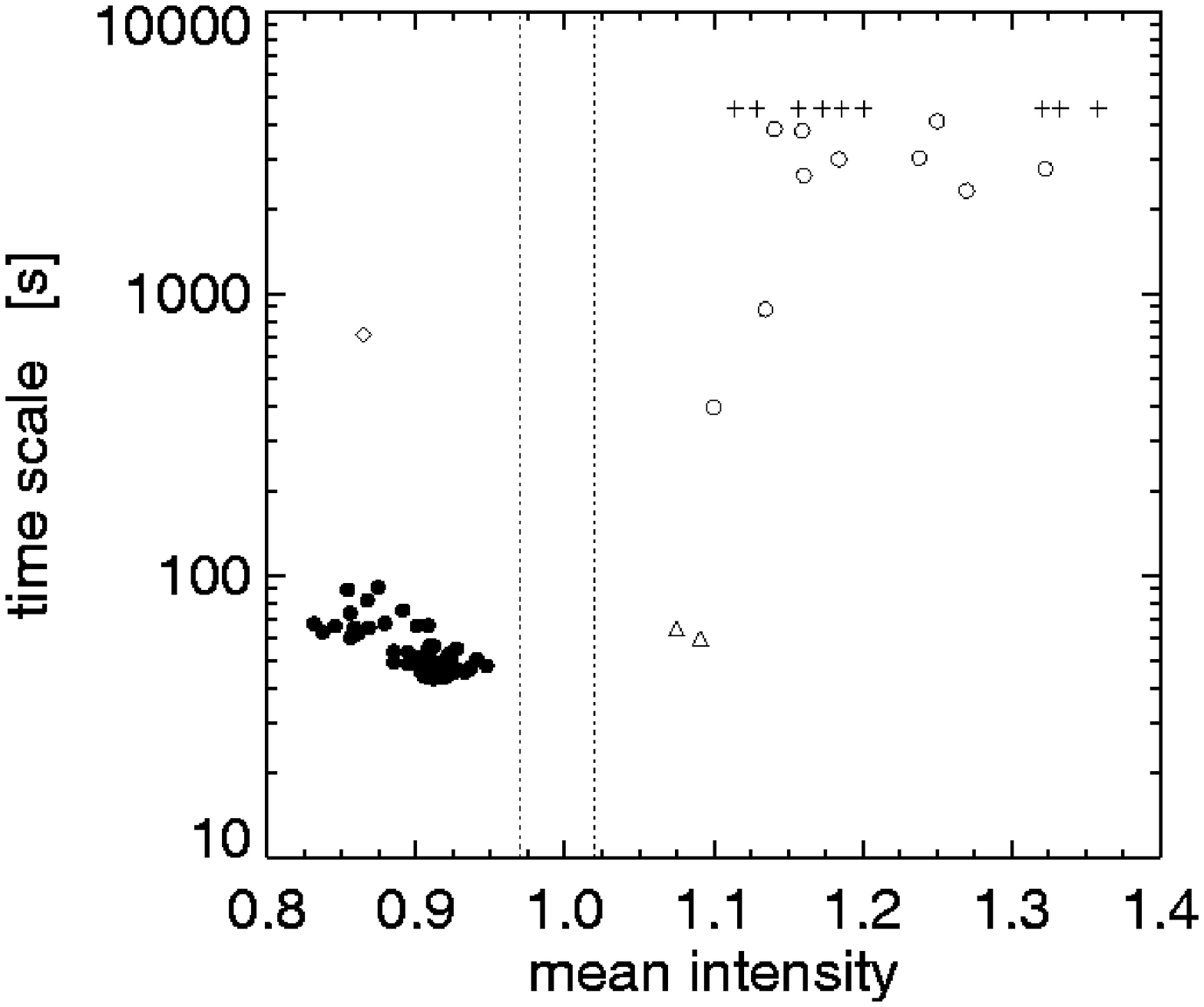}
  \includegraphics[width=3.75cm]{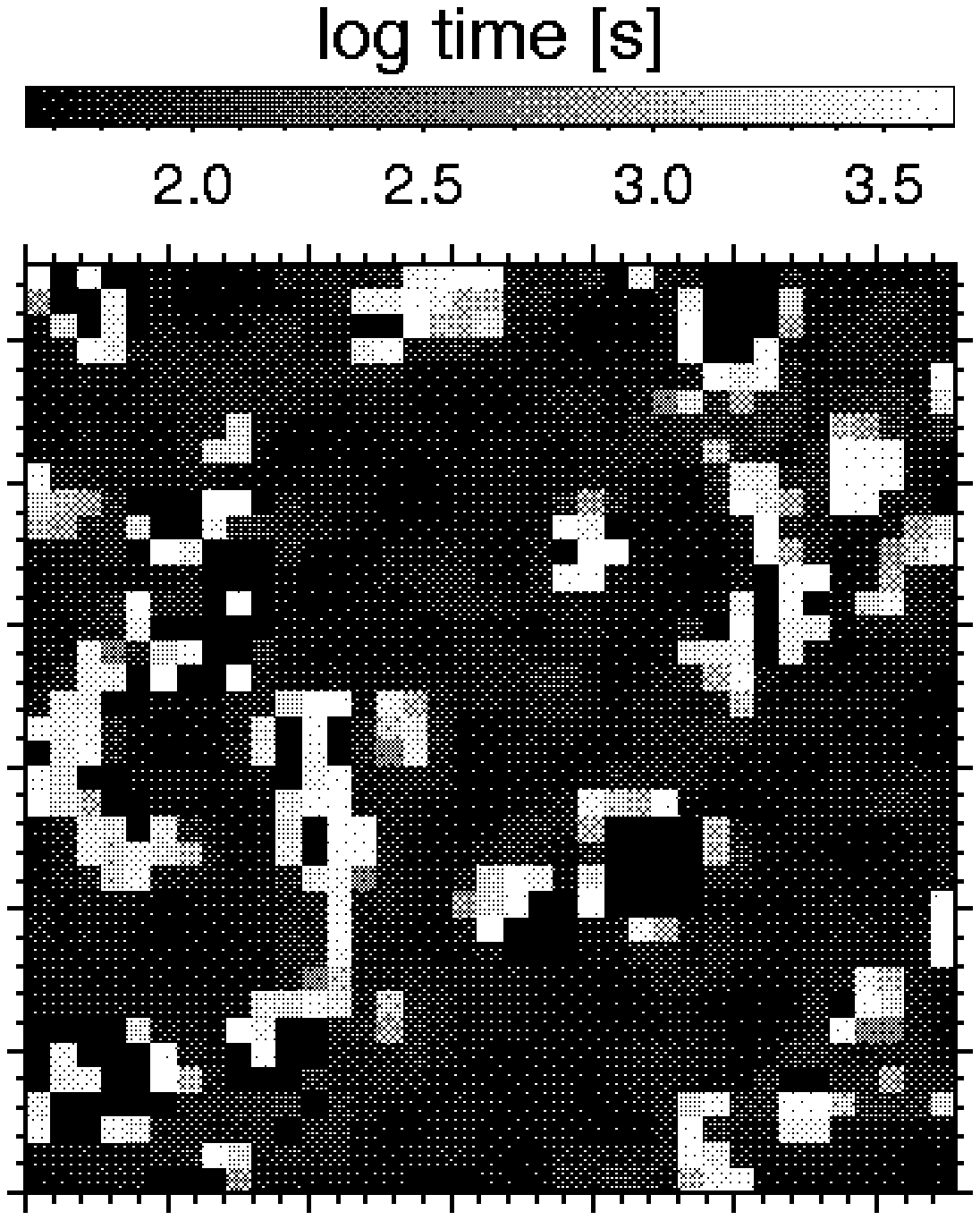}
  \caption{Left: Evolutionary time scale versus mean intensity.
    Vertical dotted lines indicate the intensity threshold chosen to
    differ between internetwork (filled circles) and network (open
    circles) subfields. Diamond: internetwork subfield with comparably
    long time scale. Triangles: network subfields with comparably
    short time scales.  Crosses: network subfields with time scales
    longer than the duration of the used data set (see text). Right:
    Evolutionary time scale per subfield.}
  \label{fig:time_scales}
\end{figure}
%

In Fig.~\ref{fig:rho} each single autocorrelation function is plotted
as a function of time for all internetwork (upper panel) and network
(lower panel) subfields, revealing a diversity in amplitude and phase.
We highlight the autocorrelation of two individual subfields in the internetwork 
and network (dash-dotted) to demonstrate that their behaviour 
stays quasi-periodic for a time lag of at least 25\,min. For both the network and the
internetwork, the first cycle in the individual and the average
autocorrelation peaks at $\sim$3\,min. The following cycles tend to
appear with a 3-4\,min period for internetwork subfields and with a
$\sim$5\,min period for network subfields.

As an evolutionary time scale we take the time after which the
autocorrelation decreased to a value of $1/e$.  For those subfields
where the autocorrelation function does not fall below a value of
$1/e$, the time scale is set to the duration of the chosen data set of
4530\,sec. Figure~\ref{fig:time_scales} shows the derived time scale
versus the mean intensity per subfield (left), and a spatial map of
all deduced time scales (right). The time scales show a clear
dichotomy between network and internetwork subfields. The internetwork time scales
group together in the range 43-91\,sec with a mean of 52\,sec.
In taking this mean we neglect one out of the 74 internetwork subfields that shows a
much larger time scale of 716\,s, most likely due to contributions of
unresolved magnetic elements. We cannot evaluate the proper mean time scales
for the network subfields because of the time constraint of $\sim$75\,min (see above). 
However, network time scales are well known to vary substantially in the range
from min to many hours \citep[e.g. ][]{liu+etal1994, muller1994, berger+title1996, 
berger+etal1998, nisenson+etal2003}.

\paragraph{Comparison with simulations.}  Evolutionary time scales
based on horizontal cross-sections of gas temperature at fixed
geometrical heights in the 3D hydrodynamical simulations of
\citet{wedemeyer+etal2004} of the quiet sun are height dependent with values in the range
20-25\,sec at chromospheric heights ($\sim 1000$\,km) and about
70\,sec in the middle photosphere \citep{wedemeyer2003}. The
temperature amplitudes of the model chromosphere might be somewhat
uncertain due to the simplified treatment (grey, LTE), whereas the
dynamics and the related timescales are connected to already
realistically modelled lower layers. An indirect qualitative
comparison can be done if one considers the following effects: (a) the
observed intensity originates from an extended formation height range
instead of a fixed geometrical height, (b) intensities refer to
corrugated surfaces of optical depth instead of plane horizontal cuts
like used for the model, (c) observations are subject to seeing
(spatial smearing) and instrumental effects, in particular spatial
resolution, (d) the simulations cover horizontally a limited range of
$\sim$6\,Mm only and do not account for dynamic phenomena larger than
these scales, (e) the simulations are purely hydrodynamic. The mean
value of 52\,sec derived from our observations is too short for being
due to reversed granulation in the middle photosphere (70\,sec in the
model) but, considering the mentioned effects, is actually in a
plausible range expected for (low) chromospheric layers.

\subsection{Image segmentation analysis}\label{ssec:patterning}

\begin{figure}[t]
  \centering
  \includegraphics[bb= 0 0 340 400, width=4cm]{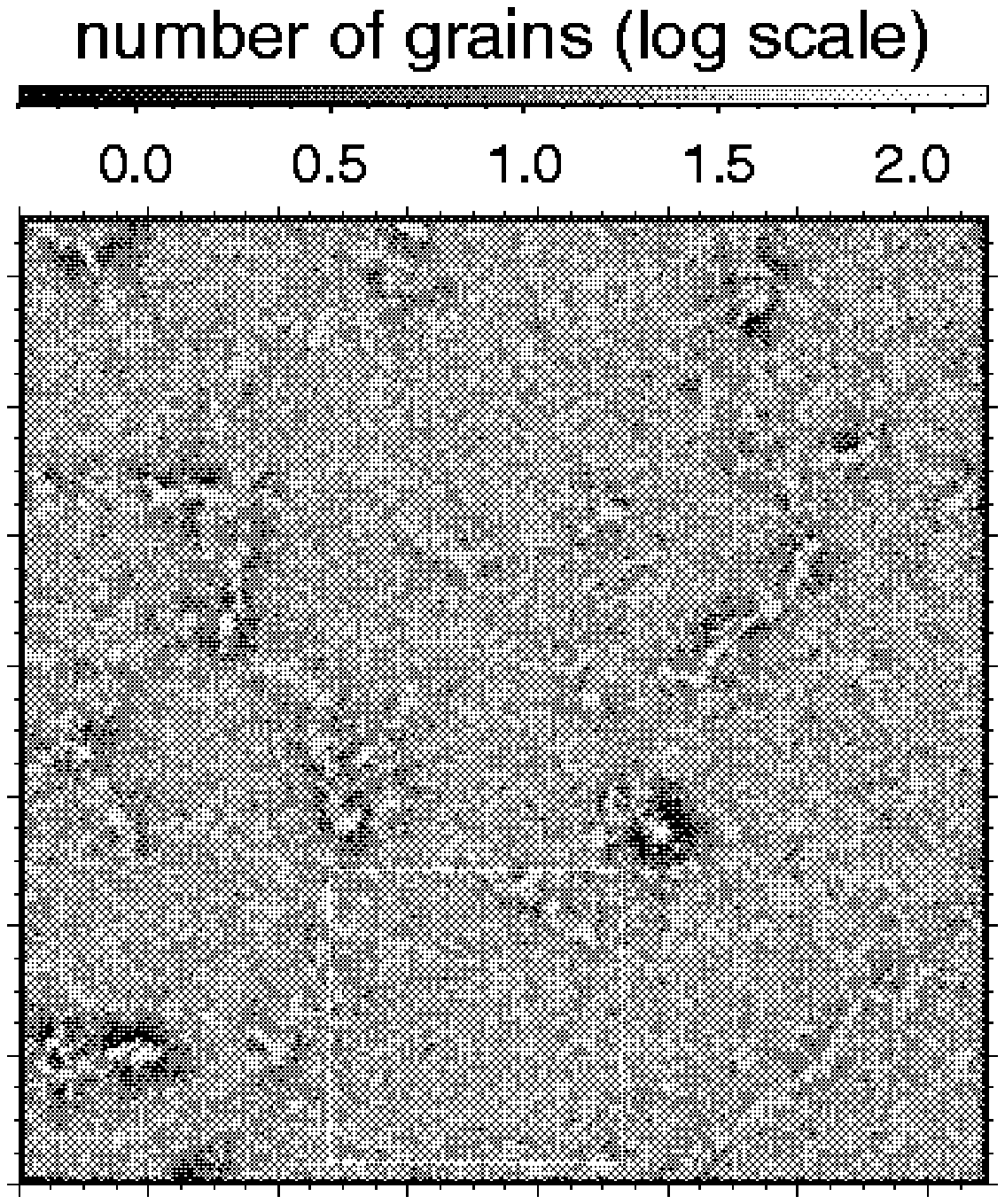}
  \includegraphics[bb= 0 0 340 400, width=4cm]{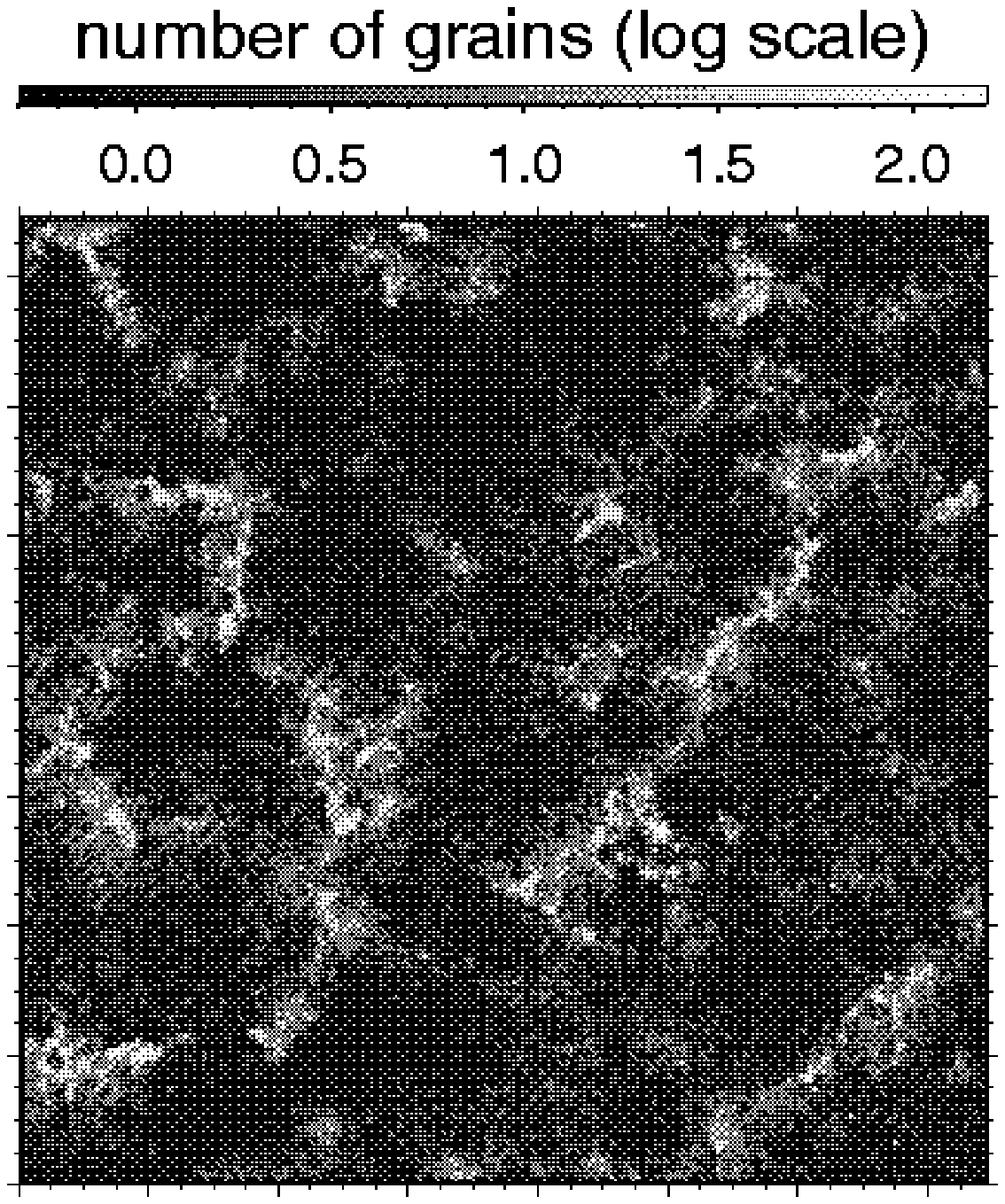}\\
  \includegraphics[width=4cm]{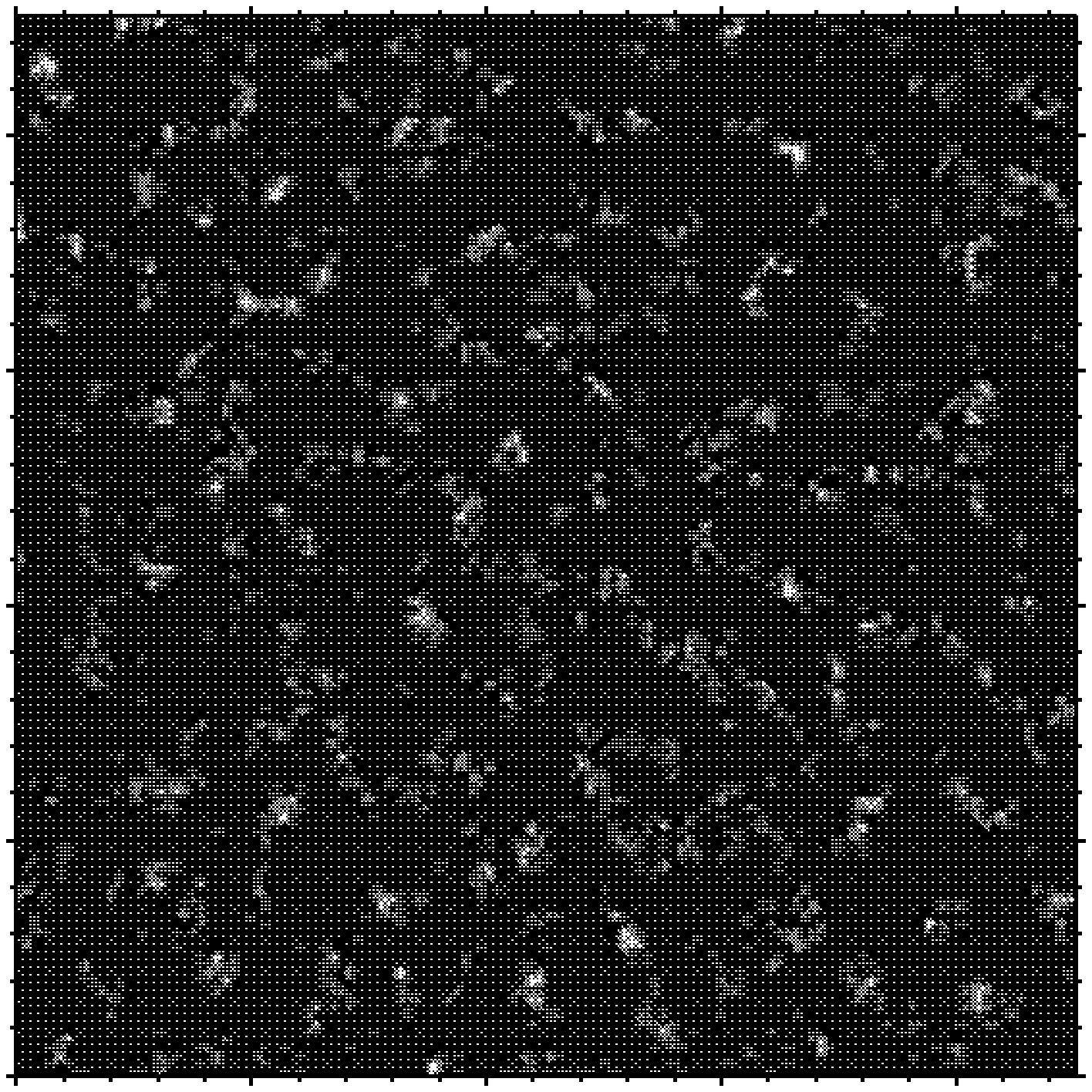}
  \includegraphics[width=4cm]{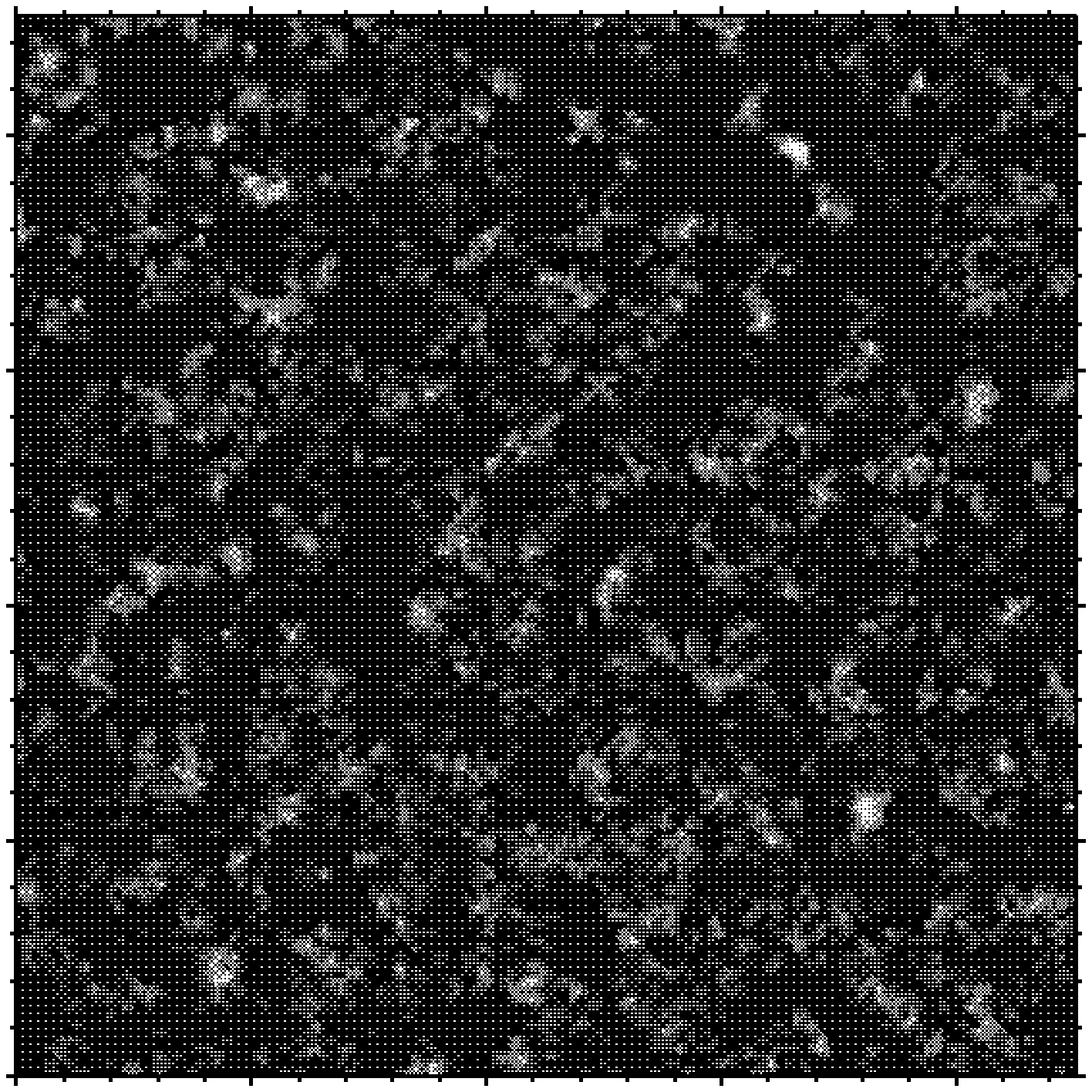}
  \caption{Upper panels: Cumulative number of bright points
           after an integration time of 350\,min (left) and with a threshold 
           intensity of $>$1.2\,$I_{\rm mean}$ (right).
           One minor tick corresponds to 5\,arcsec.
           Lower panels: Zoom-in (by a factor of two) into 
           the cumulative number of bright points 
           (at coordinates $X=48-93$\,arcsec and $Y=4-49$\,arcsec as 
           indicated by the white square in the upper panel, left) after 
           3\,min (left) and 10\,min (right) of integration. 
           One minor tick corresponds to 2\,arcsec.}
   \label{fig:grain_pos_zoom}
\end{figure}

We turn now to the question whether the internetwork brightenings
occur at preferred locations. We use the
image segmentation processing that provides a list of positions of all identified
brightenings, and compute the cumulative bright point number in
dependence of integration time. Figure \ref{fig:grain_pos_zoom} (upper panels) shows the
result for an integration time of 350\,min. We display two maps: 
if all the brightenings are taken into account (left) and only those
that have intensities $>$1.2\,$I_{\rm mean}$ (right). 
For better visibility of the internetwork regions the cumulative maps 
are displayed on a logarithmic scale. 
The top left panel of Fig.~\ref{fig:grain_pos_zoom} shows locations 
that are void of brightenings during the whole observing time. 
These are locations close-by the network and around strong magnetic elements 
where almost no brightenings appear nor move to. The map
with an intensity threshold value of 1.2\,$I_{\rm mean}$) 
(Fig.~\ref{fig:grain_pos_zoom}, upper panel, right), 
shows that there are also internetwork regions with a deficit in 
the occurrence of brightenings like the one
in the lower left part of our FOV (extending 70\,arcsec in 
x-direction and 40\,arcsec in y-direction). 
We cannot exclude completely that these internetwork regions 
directly surrounding the network are a byproduct
of the image segmentation process. Network bright points appear to be diffuse in intensity
because of scattering effects in the solar atmosphere. As a consequence localised
peaks produced by internetwork grains moving towards or emerging close by the network 
are harder to detect against the diffuse halo around the network bright points which results
in a deficit of detected brightenings. 
%
%
The lower panel of Fig.~\ref{fig:grain_pos_zoom} 
shows a zoom-in into the cumulative maps for integration times 
of 3\,min (left) and 10\,min (right).
We recognise a regular mesh-like pattern 
in the internetwork which is still clearly 
visible after an integration time of up to 
$\sim$10\,min (a typical granular life time). 
The brigthenings though appear to be mostly 
clustered on the nodes of this mesh.
Even after an integration time of 60\,min 
(not shown) the mesh is still distinctive. On average, the cell size of 
the mesh is $\sim$4-5\,arcsec. Since we mark 
the position of each brightening detected 
in every successive filtergram, 
the cumulative maps do reflect not only the 
appearance of a new internetwork grain
but also the movement of those grains 
(due to seeing and proper motion) that have been detected 
before. A movie made of 3\,min cumulative maps, 
shows evidence that the mesh pattern itself develops and migrates. 
%
\begin{figure*}[th]
  \sidecaption
  \centering
  \includegraphics[width=5.5cm]{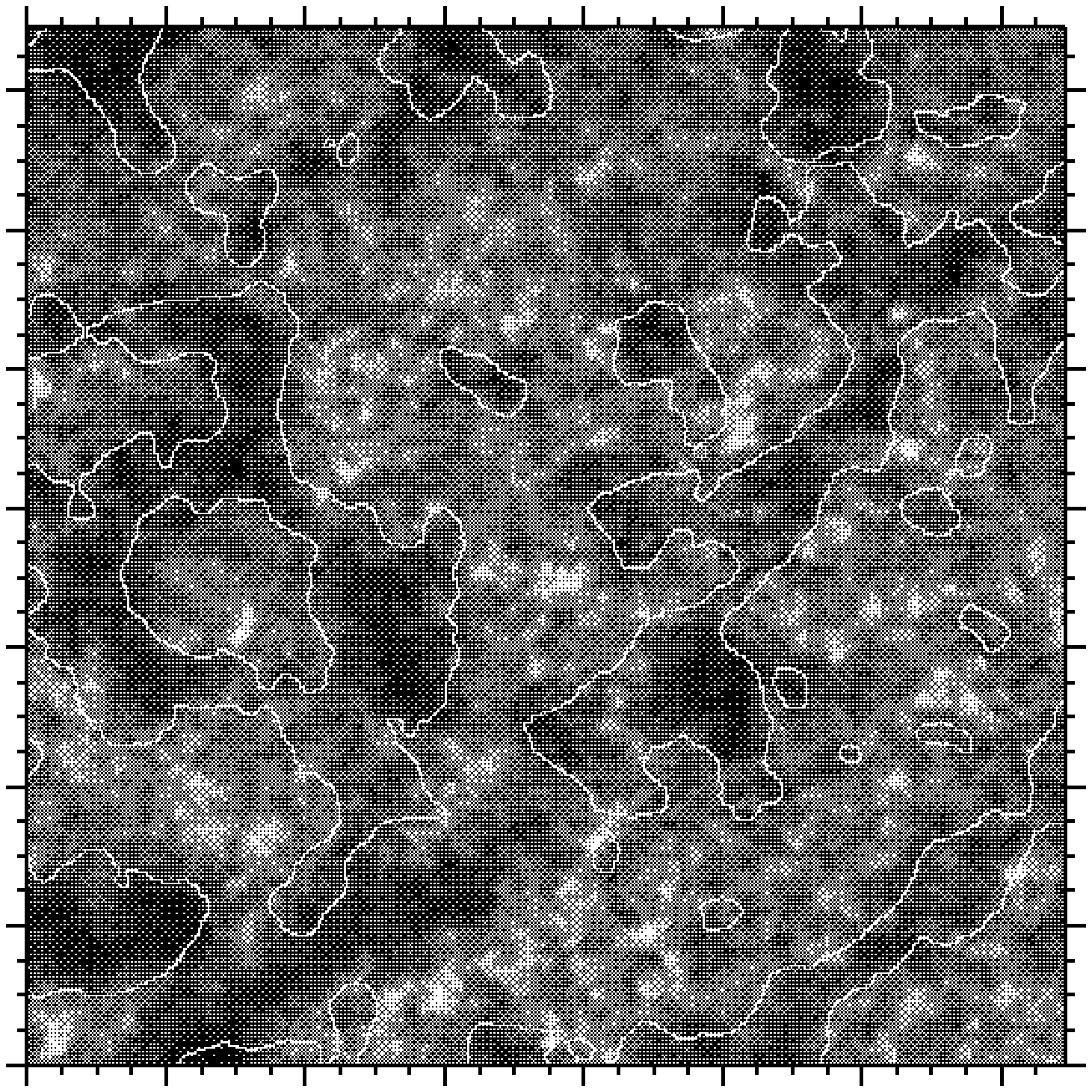}
  \includegraphics[width=5.5cm]{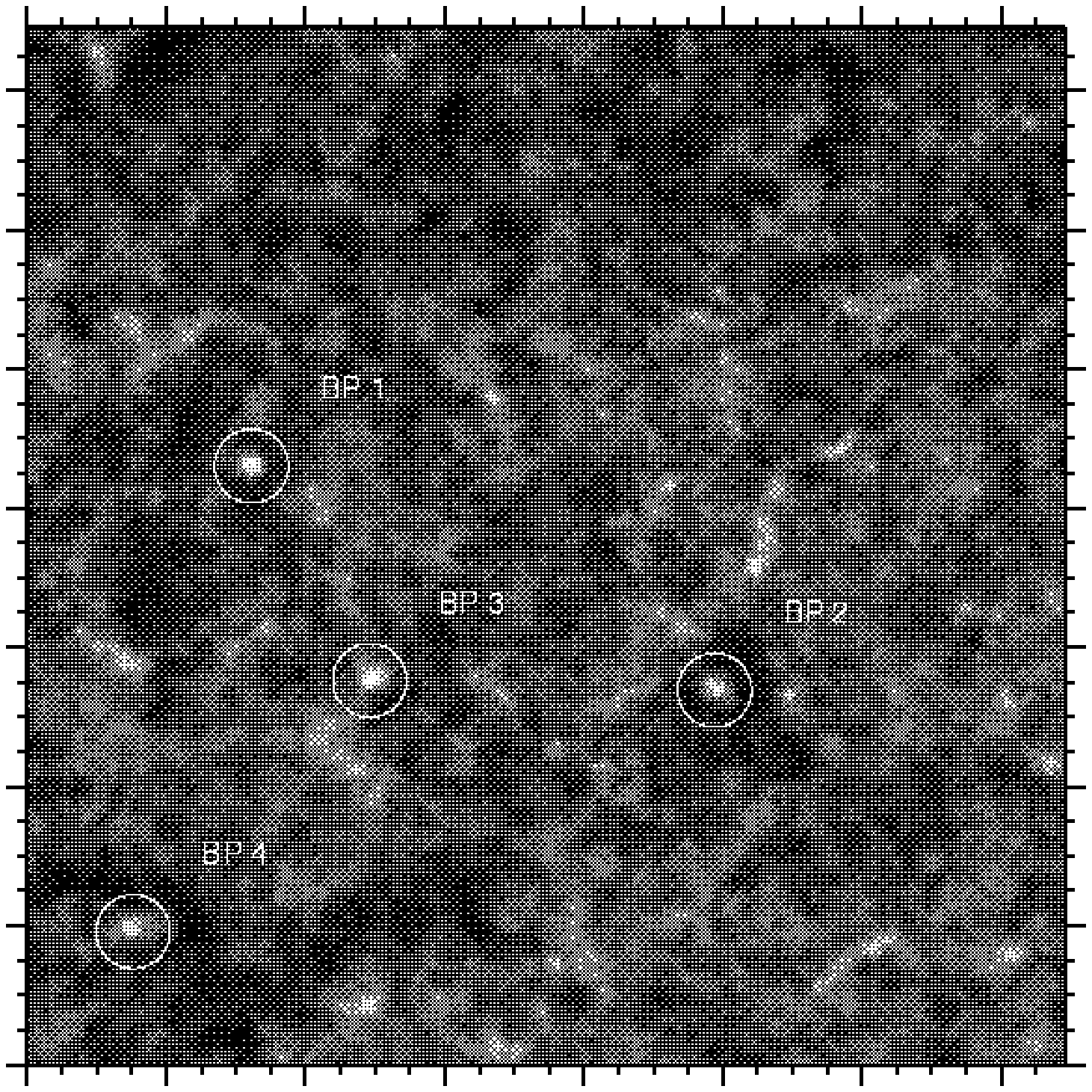}
   \caption{Power maps with 3\,min (left) and
            5\,min (right) periodicity computed over the full 6\,hours. 
            The contour lines in the 3\,min power map outline the network as determined 
            from the intensity map averaged over the whole time sequence. 
            The images are scaled independently. One minor tick mark 
            corresponds to 5\,arcsec. Circles indicate locations with 
            increased signal in the 5\,min band that are co-located
            with enhanced magnetic signal in the average MDI magnetogram 
            (see Fig.~\ref{fig:obs_sample_mdi}, c).}
   \label{fig:power_maps}
\end{figure*}
%

What is the physical mechanism that causes the bright grains to appear on
an organised mesh? Simulations of solar convection show
the organisation of strong downdraft fingers on a similar spatial scale
\citep{stein+nordlund1989}. Strong turbulence is associated with these 
downdrafts. According to \citet{rimmele+etal1995} this turbulence gives rise to acoustic noise,
which is a potential source for the \CaII~K$_{\rm 2v}$ excitation. 
This would explain that the grains appear preferentially on a meso scale dictated
by the downdraft fingers. The locations with a high number of accumulated bright points
(brightest areas in Fig.~\ref{fig:grain_pos_zoom}, lower left panel) may well
correspond to locations of such strong downdrafts. The bright points that appear in other
locations may be generated by weaker turbulence in the more pronounced
intergranular lanes that outline the meso-scale downflow pattern. The latter could
explain the slow migration of the pattern on time scales corresponding to granular
evolution while the changes in the location of the more persistent bright points
corresponds to the times over which the meso-scale downflows evolve. 
Note, however, that there is not necessarily a one-to-one correspondence between
the location of the photospheric excitation and the resulting occurrence of a bright
point \citep{hoekzema+etal1998, hoekzema+rutten1998, leenaarts+wedemeyer2005} because
the waves may not propagate vertically. 

\subsection{Power maps}\label{ssec:power_maps}

We compute the power over the $\sim$6\,hours at each pixel and apply a bandpass
filter to extract the contributions from signal with a 3\,min and
5\,min periodicity. The bandpass filters are centred at 5.5\,mHz and
3.33\,mHz, with a FWHM of 0.4\,mHz each. Large-scale intensity variations 
in time are eliminated by a linear fit to the data points.
To apodise we apply a one-dimensional Hanning window function that
covers 8\,\% of the data vector length and tapers the edges to zero. 

Figure \ref{fig:power_maps} shows the resulting power maps thus visualising the
well-established effect that internetwork and network respond in a different way to
wave motions most likely excited in the photosphere \citep{liu+sheeley1971, 
cram+dame1983, dame+gouttebroze+malherbe1984, lites+rutten+kalkofen1993a, 
lites+rutten+thomas1994, judge+tarbell+wilhelm2001}. 
The network participates predominantly in long-period oscillations ($>$5\,min)
while in the internetwork short-period oscillations ($\sim$3\,min, and shorter) 
dominate the picture. 

From the 3\,min map we recover that there are regions
that according to our definition clearly belong to the internetwork, 
but show a significant suppression of 3\,min power. The most 
conspicuous area is the slanted one between $X=20-70$\,arcsec and $Y=0-35$\,arcsec
(also apparent in Figures \ref{fig:avg_hist_mask}, left, and \ref{fig:grain_pos_zoom}, 
upper panel, left). These voids have been observed before 
\citep[e.g. ][]{judge+tarbell+wilhelm2001, krijger+etal2001}
and named {\it magnetic shadows} by \citet{judge+tarbell+wilhelm2001}.
The phenomenon is ascribed to the spreading of the magnetic field with height and 
forming the canopy, shadowing its immediate 
surroundings and thus hampering (or modifying) upward propagating waves
\citep[see also ][]{rosenthal+etal2002, bogdan+etal2003}.
The shadows also appear around strong
network bright points with a persistent magnetic signal as can be seen in the 
right panel of Fig.~\ref{fig:power_maps}
and in addition as narrow lanes along the network as indicated by the 
map of the cumulative number of bright points 
(see Fig.~\ref{fig:grain_pos_zoom}, upper panel, left). 
Inspecting the radial behaviour of azimuthally 
averaged power spectra from the centre of the indicated network bright points
(BP1-BP4 in Fig.~\ref{fig:power_maps}, right) outwards 
we find the tendency that the integrated power is sharply 
decreasing with distance giving rise to the shadows around the bright points.
Power increases again to average levels in the internetwork 
beyond the indicated circles. 

%
%

\section{Morphology}\label{sec:statistics}

The results of image segmentation are visualised in terms of occurrence distributions 
separately for the internetwork and the network. 
Our network and internetwork masks (Sect.~\ref{ssec:masks})
avoid regions that show mean intensities between $0.97<I_{\rm mean}<1.02$ by definition.
For comparison, we also show the occurrence distribution 
representative for all identified brightenings inside the FOV. 
Note that these occurrence distributions include 
brightenings with $0.97<I_{\rm mean}<1.02$. 
Each occurrence distribution is normalised to 
the total number of identified brightenings. 
%
\begin{figure*}[th]
  \centering
  \includegraphics[width=5.75cm]{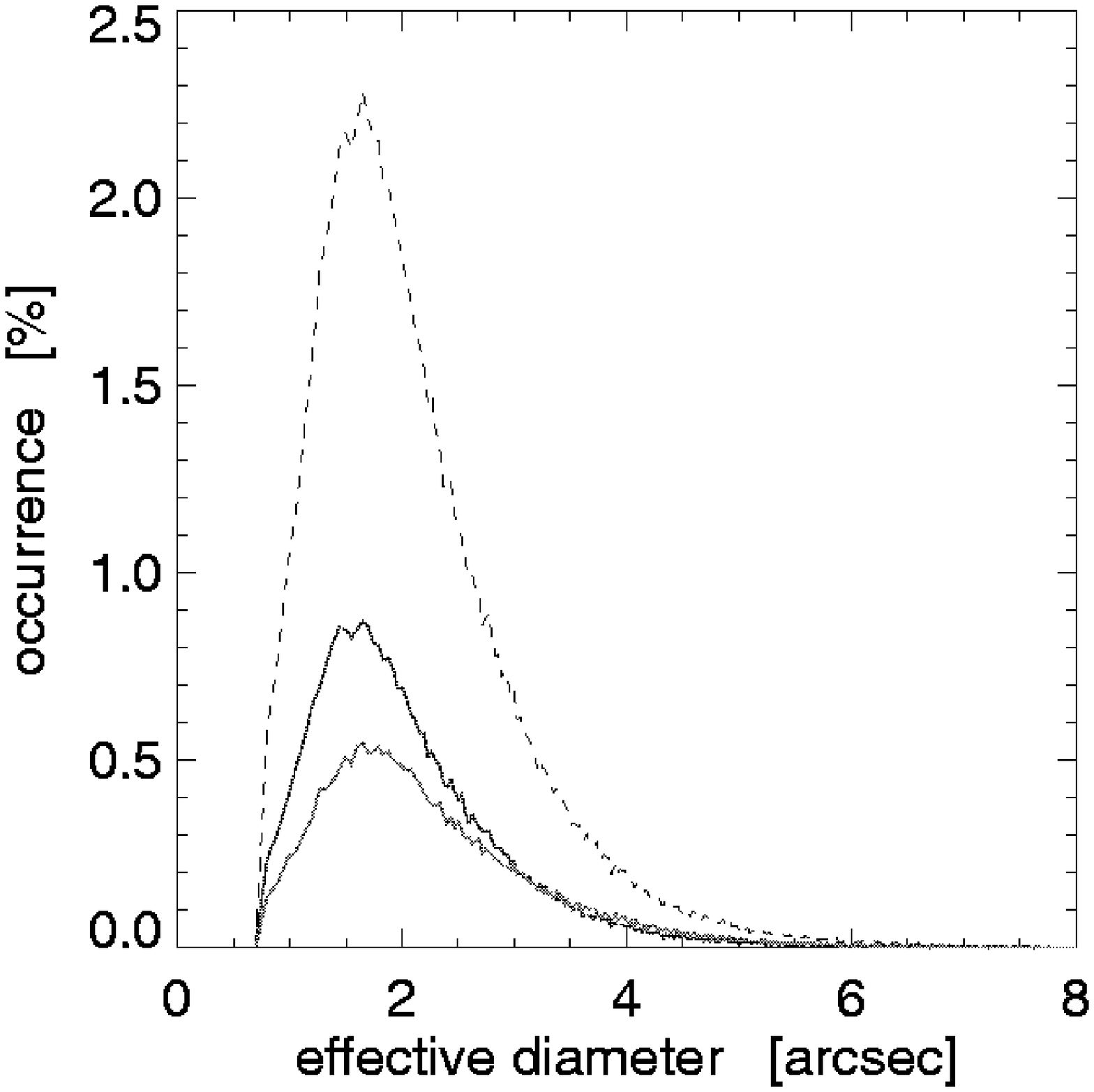}
  \includegraphics[width=5.75cm]{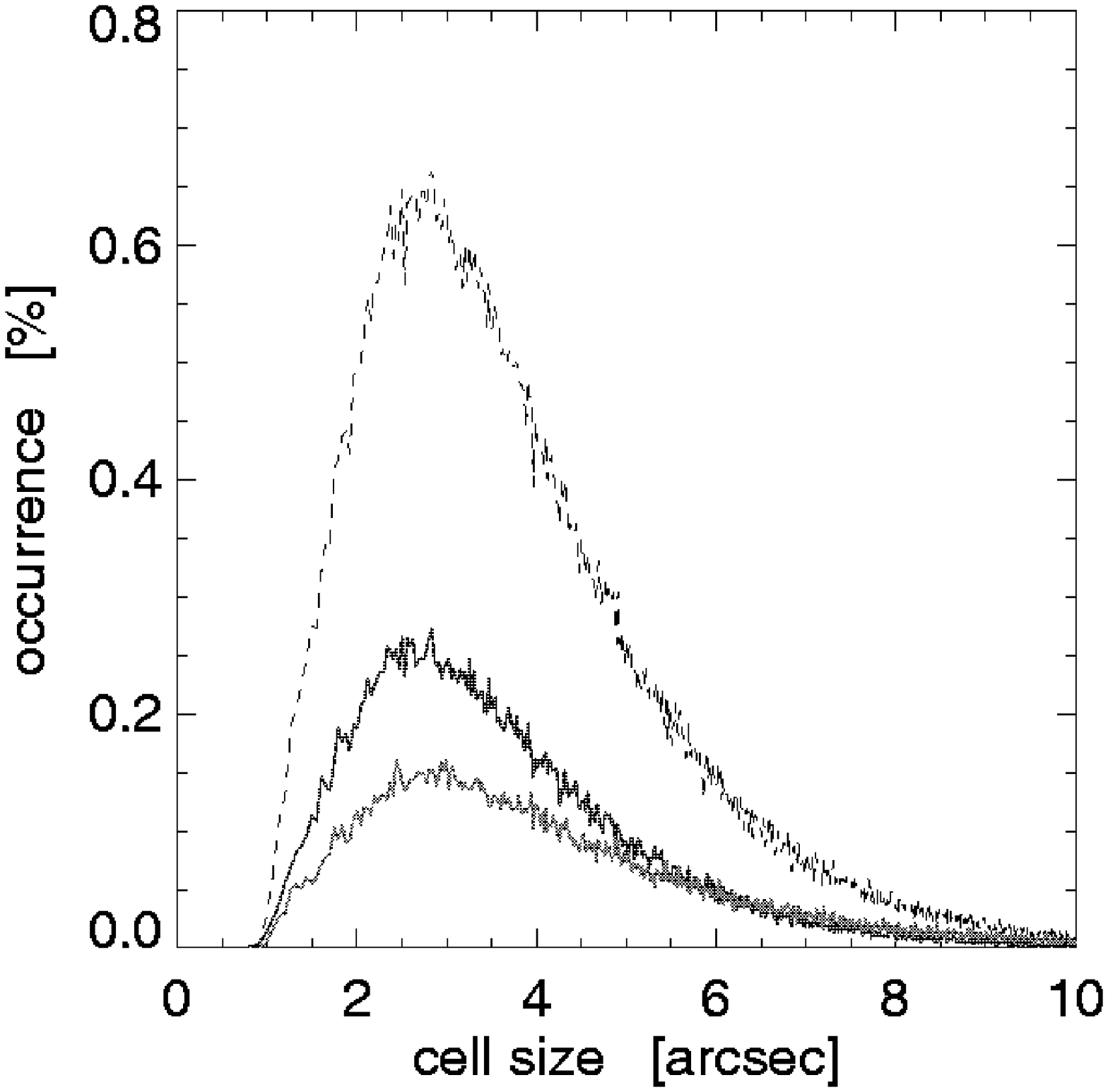}
  \includegraphics[width=5.75cm]{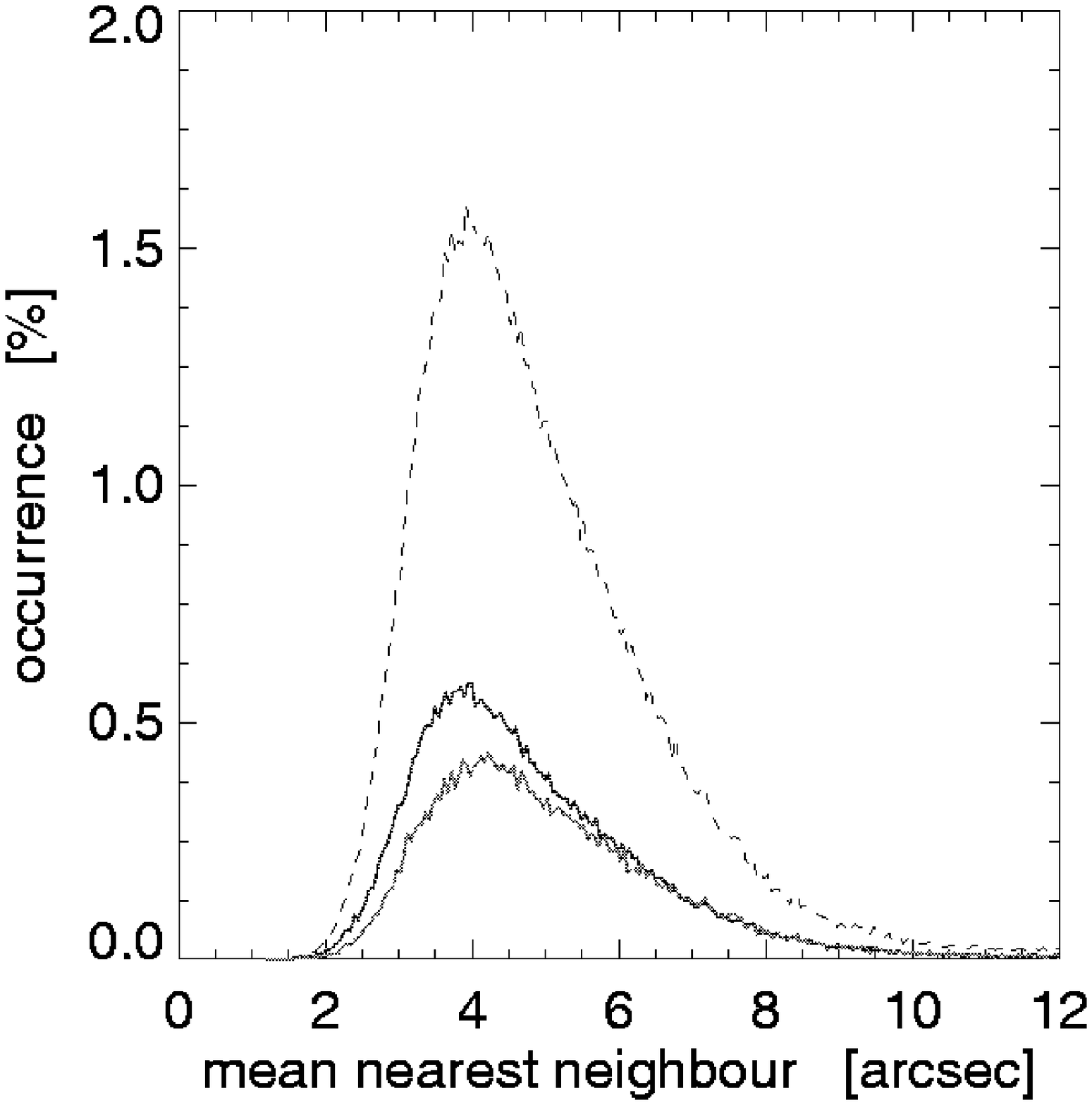}
  \caption{Occurrence distributions of the effective diameter (left), the cell size (middle),
           and the nearest neighbour distance (right). The distributions are normalised
           with respect to the corresponding number of identified brightenings. For each quantity
           the distributions for the network (grey), the internetwork (black), and 
           all identified structures (dashed) are displayed.}
  \label{fig:sizes}
\end{figure*}
%

Since we do not track each individual identified brightening in time, our sample of
brightenings is not statistically independent. Therefore, and 
to be more independent from seeing conditions, 
only those filtergrams contribute to the distributions, that show a 
rms-contrast of 7.5\,\% higher than the value of a 
smoothed version of the contrast variation at the same
time step (see Fig.~\ref{fig:rms07}). This leads to a selection of 
360 filtergrams with a mean cadence of 57\,sec (not equidistant). 
We count only those structures that span more than two pixels in both cell size
and effective diameter, and that have nearest-neighbour distances larger than 3 pixels. 
This leaves us with a sample of 206056 identified structures.

\subsection{Size distributions}\label{ssec:sizes}

The image segmentation techniques (see Sect.~\ref{sec:imgseg}) 
are used to derive three different size measures: effective diameters, cell sizes, 
and mean distances to next neighbour elements. Figure \ref{fig:sizes}
displays the resulting occurrence distributions. They are 
single-peaked and asymmetric with 
a steep rise to the peak, followed by a gradual fall-off with size.
All size measures peak almost at equal 
values for the network and the internetwork: we find typical effective diameters of
1.6\,arcsec for both internetwork and network structures. We partly ascribe 
this to the limitations of the image segmentation process.
Characteristic cell sizes are almost twice as large as effective diameters and peak 
around 2.8\,arcsec and 3.0\,arcsec while nearest-neighbour distances 
are typically 4.0\,arcsec and 4.2\,arcsec for the
internetwork and network, respectively. Arithmetic mean values are 
larger then the characteristic values since they reflect 
the asymmetry of the corresponding distribution function. 
The mean value for the nearest-neighbour distance in the internetwork amounts to 
5\,arcsec quantifying and confirming the impression we got from the 
the cumulative bright point number maps (see Fig.~\ref{fig:grain_pos_zoom}) 
that reveal the existence of a $\sim$4-5\,arcsec (judged by eye) 
mesh-like pattern in the internetwork. 

\subsection{Discussion}

The simulations of \citet{carlsson+stein1997} do reproduce
the spectral and temporal behaviour of the \CaII~H line
profile quite well but are one-dimensional and thus cannot predict
morphological properties. On the other hand, current 
three-dimensional radiative hydrodynamic simulations
\citep{wedemeyer+etal2004} have to neglect time-dependent
Hydrogen ionisation and non-grey NLTE radiative transfer, so
that no realistic \CaII~K line profiles can be synthesised.

Seeing conditions, differential image motion, 
and instrumental effects influence the spatial resolution and thus affect 
the image segmentation algorithm. We resolve structures that are larger than 0.7\,arcsec. 
Spatially unresolved conglomerates of small structures
are interpreted as a large single structure, which leads to an underestimation 
of the frequency of small-scale brightenings. We argue that this is predominantly
the case for the internetwork. It is most likely that we predominantly  
detect the brightest structures and are unable to resolve their substructure.
Underestimation of the occurrence of faint structures
on the one side and not resolving the substructure of the rare brightest structures
shifts e.g. the peak of the size distributions to larger scales.
This might explain the similarities found in the size distribution 
functions of the network and the internetwork. To clarify this data sets with 
better spatial resolution are needed.

What is the physical mechanism that causes the bright grains to appear on
an organised mesh (see Sect.~\ref{ssec:patterning})? Simulations of solar convection show
the organisation of strong downdraft fingers on a similar spatial scale
\citep{stein+nordlund1989}. Strong turbulence is associated with these 
downdrafts. According to \citet{rimmele+etal1995} this turbulence gives rise to acoustic noise,
which is a potential source for the \CaII~K$_{\rm 2v}$ excitation. 
This would explain that the grains may appear preferentially on a meso scale dictated
by the downdraft fingers. The locations with a high number of accumulated bright points
(brightest areas in Fig.~\ref{fig:grain_pos_zoom}, lower left panel) may well
correspond to locations of such strong downdrafts. The bright points that appear in other
locations may be generated by weaker turbulence in the more pronounced
intergranular lanes that outline the meso-scale downflow pattern. The latter could
explain the slow migration of the pattern on time scales corresponding to granular
evolution while the changes in the location of the more persistent bright points
corresponds to the times over which the meso-scale downflows evolve. 
Note, however, that there is not necessarily a one-to-one correspondence between
the location of the photospheric excitation and the resulting occurrence of a bright
point \citep{hoekzema+etal1998, hoekzema+rutten1998, leenaarts+wedemeyer2005} because
the waves may not propagate vertically. 

%
%

\section{Conclusions}\label{sec:conclusions}

We analysed a long time sequence of narrowband \CaII~K filtergrams 
to investigate the morphological, dynamical, and evolutionary,
properties of the solar chromosphere separately for the internetwork
and the network. Quantifying these properties is important for 
comparison with current and future numerical simulations.

From an autocorrelation analysis on small subfields 
we find average evolutionary time scales of 52\,sec in the internetwork. 
Individual values can be found in the range 40-90\,sec. 
The predicted time scales of temperature variations 
in the numerical simulations compare favourably
with the observed time scales of intensity variations 
in the internetwork notwithstanding the severe 
simplifications in the simulations.
Differences between the two time scales 
may occur because of intensity variations are the result of variations
of temperatures over a range of optical depths and these vary with height
themselves. In addition, seeing and instrumentation affect the spatial 
resolution leading to longer time scales. 

To identify and isolate small-scale features we apply image segmentation 
techniques and find evidence for internetwork
grain patterning on spatial scales of $\sim$4-5\,arcsec (see
Sect.~\ref{ssec:patterning}, Fig.~\ref{fig:grain_pos_zoom}). 
We argue that this pattern is related to photospheric downdrafts,
which occur on similar spatial scales, through the enhanced excitation
of acoustic waves that these downdrafts provide. 

We separated network from the internetwork by a double-Gaussian decomposition 
of the intensity distribution averaged over the full duration of the observation.
The validity of this decomposition is confirmed by the 
3\,min power map (Fig.~\ref{fig:power_maps}, left) in which the power deficit on the network 
coincides almost perfectly with the network mask. There is, however, one region (lower left)
that appears dark in both the mean intensity map (Fig.~\ref{fig:obs_sample_mdi}, left)
and the 3\,min power map. We speculate that this region contains magnetic field
that has not penetrated yet into the chromosphere. Either the magnetic field
hinders the propagation of acoustic waves or these waves suffer mode conversion at
low altitude before giving rise to \CaII~K emission. The latter is also the mechanism
proposed for the small-scale shadows we see around strong network 
bright points (Fig.~\ref{fig:power_maps}, right).

From a statistical analysis based on results from image segmentation 
we find that internetwork grains and network bright points 
appear to have very similar typical sizes ($\sim$1.6\,arcsec). This is somewhat surprising because
of their different physical nature as is evident from differences in
the power spectra and the disparity in their temporal behaviour.
We surmise that this has to be attributed in part to insufficient spatial resolution. 

For better understanding of the statistical properties in the \CaII~K filtergrams 
and how they are related to physical properties in the solar atmosphere both simulations
that allow realistic modelling the formation of \CaII~K, and observations with 
higher spatial and spectral resolution are required, preferably in the form of
two-dimensional spectroscopy.

%
%

\acknowledgements

The VTT is operated by the Kiepenheuer--Institut
f\"ur Sonnenphysik, Freiburg, Germany. Part of this work (AT) 
was supported by the DFG under grant, by NSF under grants ATM 00-86999 and 
AST MRI 00-79482, and by NASA under grant NAG 5-12782. We thank the referee
Rob Rutten for extensive comments. 

%
%


%
\end{document}